\documentclass[12pt]{article}

\def\BibTeX{{\rm B\kern-.05em{\sc i\kern-.025em b}\kern-.08em
    T\kern-.1667em\lower.7ex\hbox{E}\kern-.125emX}}
\usepackage{times,amsmath,epsfig}
\usepackage{amssymb}
\usepackage{amsthm}
\usepackage{epstopdf}
\usepackage{cite}
\usepackage{color}
\usepackage{dcolumn}
\usepackage{array,multirow}

\usepackage[text={6.7in,8.8in},centering]{geometry}

\usepackage{tabularx}
\usepackage{psfrag}

\usepackage{graphicx}

\newtheorem{rem}{Remark} 

\begin{document}
\begin{flushleft}

\textbf{\Large Title}\quad{\large Adaptive Sparse-grid Gauss-Hermite Filter}\\[20pt]

\textbf{\large Authors}\quad{Abhinoy Kumar Singh$^{*}$,  Rahul Radhakrishnan, Shovan Bhaumik and Paresh Date }\\[10pt]
\textbf{Institutional affiliations of authors}       \\[15pt]
\textbf{$^{*}$Corresponding author}\\
Abhinoy Kumar Singh, \\
Department of Electrical Engineering,\\
Indian Institute of Technology Patna, Bihar-801103, India\\
E-mail: abhinoy@iitp.ac.in\\[15pt]

Rahul Radhakrishnan, \\
Department of Electrical Engineering,\\
Indian Institute of Technology Patna, Bihar-801103, India\\
E-mail: rahul.pee13@iitp.ac.in\\[15pt]

Dr. Shovan Bhaumik,\\
Department of Electrical Engineering,\\ 
Indian Institute of Technology Patna, Bihar-801103, India\\
Phone +91 612 255 2049, Fax: +91 612 2277383 \\
E-mail: shovan.bhaumik@iitp.ac.in\\[15pt]

 Dr. Paresh Date, \\
Department of Mathematics, \\
College of Engineering, Design and Physical Sciences,\\ 
 Brunel University, 
 London
Uxbridge, UB8 3PH, United Kingdom\\
Phone: +44 1895 265613, Fax:     +44 1895 269732 \\
Email: Paresh.Date@brunel.ac.uk \\[15pt]

\end{flushleft}

\newpage

\begin{abstract}
In this paper, a new nonlinear filter based on sparse-grid quadrature method has been proposed. The proposed filter is named as adaptive sparse-grid Gauss-Hermite filter (ASGHF). Ordinary sparse-grid technique treats all the dimensions equally, whereas the ASGHF assigns a fewer number of points along the dimensions with lower nonlinearity. It uses adaptive tensor product to construct multidimensional points until a predefined error tolerance level is reached. The performance of the proposed filter is illustrated with two nonlinear filtering problems. Simulation results demonstrate that the new algorithm achieves a similar accuracy as compared to sparse-grid Gauss-Hermite filter (SGHF) and Gauss-Hermite filter (GHF) with a considerable reduction in computational load. Further, in the conventional GHF and SGHF, any increase in the accuracy level may result in an unacceptably high increase in the computational burden. However, in ASGHF, a little increase in estimation accuracy is possible with a limited increase in computational burden by varying the error tolerance level and the error weighting parameter. This enables the online estimator to operate near full efficiency with a predefined computational budget.
\end{abstract}

\begin{flushleft}
\end{flushleft}
\newpage

\section{Introduction}
In this article, we address the state estimation problem of a discrete nonlinear dynamic system with additive noise. The process and measurement model of the nonlinear system can respectively be defined as
\begin{equation}\label{eq3}
\textbf{x}_{k}=\phi(\textbf{x}_{k-1})+w_{k}
\end{equation}
and
\begin{equation}\label{eq4}
y_{k}=\gamma(\textbf{x}_{k})+v_{k},
\end{equation}
where $\textbf{x}_{k} \in \mathbb{R}^{n}$ represents the unknown states of the system, $y_{k}\in \mathbb{R}^{p}$ denotes the measurement at any discrete time $k$. $\phi(\cdot)$ and $\gamma(\cdot)$ are known nonlinear functions. The process and measurement noises are represented by	$w_{k} \in \mathbb{R}^{n} $ and $v_{k} \in \mathbb{R}^{p} $ respectively. They are assumed to be uncorrelated and normally distributed with zero mean and covariance, $Q$ and $R$ respectively. 

Bayesian estimation framework is a widely employed method for addressing a filtering problem. In this framework, by using the measurement likelihood and the predicted motion of the unknown states, the posterior probability density functions (pdf) are computed \cite{bar2004estimation}. 

During filtering of nonlinear systems, a set of intractable integrals appear and hence no optimal solution exists. In a widely accepted approach, the conditional pdfs are approximated as Gaussian and characterized with mean and covariance. Under this approach, a variety of filters like extended Kalman filter (EKF) \cite{bar2004estimation}, unscented Kalman filter (UKF) \cite{julier1997new} and its extensions \cite{ukfaa,ukfaa1}, cubature Kalman filter (CKF) \cite{arasaratnam2009cubature} and its extension \cite{ckfaa}, central difference filter (CDF) \cite{liu2011nonlinear} \textit{etc.} are proposed. In a different approach, particle filter (PF) \cite{arulampalam2002tutorial} is developed which approximates the true probability density function (pdf) with the help of particles and their assigned weights. Although the particle filter has high accuracy, its high computational burden restricts applicability in real time applications.

To achieve a higher accuracy under assigned computational budget, another Gaussian filter named Gauss-Hermite filter (GHF) \cite{arasaratnam2007discrete} was introduced. GHF makes use of Gauss-Hermite quadrature rule for univariate systems. This univariate quadrature rule is extended to multidimensional domain by using the product rule, which in turn results in an exponential rise in multivariate quadrature points and hence suffers from the \textit{curse of dimensionality} problem. This hinders the practical applicability of the filter for higher dimensional problems. We focus our study on decreasing the computational load of Gauss-Hermite filter without hampering its accuracy.


In an earlier approach, sparse-grid Gauss-Hermite filter (SGHF) was introduced which achieves similar accuracy as compared to the GHF, with reduced computational load \cite{jia2012sparse}. In this technique, the univariate quadrature rule is extended to multivariate case with the help of Smolyak rule \cite{smolyak1963quadrature,gerstner1998numerical}. 

In this paper, we propose a novel approach which further reduces the computational burden of Gauss-Hermite filtering. The proposed method is named as adaptive sparse-grid Gauss-Hermite filter (ASGHF). The conventional sparse-grid method treats all the dimensions equally, by default, resulting in no immediate advantage for problems where the dimensions are of differing nonlinearity.                                                                                                                                                                                                                                                                                                                                                                                                                                                                                                                                                                                                                                                                                                                                                                                                                                                                                                                                                                                                                                                                                                                                                                                                                                                                                                                                                                                                                                                           But the proposed method uses adaptive sparse-grid technique \cite{gerstner2003dimension}  which automatically finds the dimensions with comparatively lower degree of nonlinearity and generate fewer points for approximation along them which further results in reduced computational cost.

Another advantage of using this method is that it provides a smooth relation between accuracy and computational burden. Unlike the GHF and the SGHF, a small rise in computational burden is possible in the proposed method for a corresponding small increase in the accuracy, by varying the predefined tolerance level and error weighting parameters. It enables the system to work with maximum efficiency possible within the allotted computational budget.

\section{Sparse-grid Gauss-Hermite filter}\label{sec_SGHF}
While computing the mean and covariance matrix in an approximate Gaussian filter such as the GHF or SGHF, one encounters integrals of the form: 
\begin{equation}\label{eq12}
\textbf{I}_{n}(f^n(\textbf{x}))=\int_{R^{n}}f^n(\textbf{x})\mathcal{N}(\textbf{x};0,\texttt{I}_n)d\textbf{x},
\end{equation}
where $f^n(\textbf{x})$ is an $n$-dimensional nonlinear function and $\texttt{I}_n$ is an $n$-dimensional unity matrix. In SGHF, this integral is approximated using Smolyak rule which makes use of difference formulas $\bigtriangleup_{l}f^{1}(\textbf{x})=(I_{l}-I_{l-1})f^1(\textbf{x})$; $I_{0}=0$. Here $I_{l}$ is a single dimensional quadrature rule with $(2l-1)$ univariate quadrature points. The set of points and weights for $I_{l}$ can be generated using any of the moment matching method and Golub's Technique \cite{arasaratnam2007discrete}. Using Smolyak rule \cite{gerstner2003dimension},
 \begin{equation}\label{eq1}
 \begin{split}
\textbf{I}_{n}(f^n(\textbf{x}))
&=\sum_{\vert \mathbb{I} \vert_{n,L} \leq L+n-1} (\bigtriangleup_{l_{1}} \otimes \cdots \otimes \bigtriangleup_{l_{n}})f^n(\textbf{x})\\
&=\sum_{\Xi \in N^{n}_{q}} (\bigtriangleup_{l_{1}} \otimes \cdots \otimes \bigtriangleup_{l_{n}})f^n(\textbf{x}),
\end{split}
 \end{equation}
 where $\vert \mathbb{I} \vert_{n,L}$ represents an $n$ dimensional index set with accuracy level $L$ and $\otimes$ stands for tensor product. $\Xi = \begin{matrix} [l_1 & l_2 \cdots l_n \end{matrix}]^T$ represent a vector and $N_q^n$ is defined as
 \begin{align*}
\begin{split}
N_q^n&=\left\lbrace\Xi:\sum_{j=1}^{n}l_j=n+q \right\rbrace \quad \quad \quad \text{for}\quad q\geq0 \\
&=\varnothing \quad \quad \quad \quad \quad \quad \quad \quad \quad \quad\quad \quad\text{for}\quad q<0
\end{split}
\end{align*}
where $\varnothing$ is null set and $q$ is an integer \textit{i.e.} $L-n\leq q\leq L-1$.

\section{Adaptive sparse-grid Gauss-Hermite filter}\label{sec_ASGHF}


As discussed earlier, SGHF reduces the computational load of GHF. But still it suffers from two disadvantages:
\begin{enumerate}
\item Although the computational burden of the SGHF is lower than the GHF, it rises sharply with dimension of the system.

\item The accuracy level $L$ is the only parameter to control the accuracy versus computational burden relation for SGHF. Even a unit increase in it often results in drastic increase in computational burden. Due to this, the online system usually works below its efficiency under the assigned computational budget.
\end{enumerate}

To overcome the above mentioned shortcomings, we propose a modification to SGHF which is named as adaptive sparse-grid Gauss-Hermite filter (ASGHF). The proposed method reduces the computational burden of SGHF without compromising with accuracy. Further, the accuracy and the computational burden can be controlled using two predefined parameters, namely the local error indicator and the error tolerance, which are discussed in subsequent  part of this section. This provides a far better tuning control on the accuracy versus computational burden relation than the use of accuracy level in SGHF. Hence the online system can be made to work near its full efficiency.

\subsection{Notation} 


\begin{enumerate}

\item Index set $\mathbb{I}_n$: Let $Z^\phi$ with $\phi = 1, 2, \cdots$ denotes subsets of $\mathbb{Z}_+$ (set of all positive integers),  each of which may or may not be finite. Then, an index set of dimension $n$ can be defined as $ \mathbb{I}_n^\phi = \lbrace \lambda : \lambda = ( \lambda_1 \;  \lambda_2 \; \cdots \; \lambda_n ), \; \lambda_i \in Z^\phi \rbrace. $ A possible example of an index set for $n = 2$ is ${ (1  \;1), (2  \;1), (1 \; 2), (2  \;2), (3 \;1), \cdots }$. Since the order of subsets $Z^\phi$ is of no relevance to the subsequent discussion, we drop the superscript $\phi$ henceforth; keeping in mind that each $\mathbb{I}_n$ has a different set of positive integer valued vectors in general. Moreover, one can notice here that the index set $\mathbb{I}_n$ is an ordered set.
 
 \item Forward index: A set of forward indices for each fixed vector $\lambda$ is given by $\lambda + e_j$, $j = 1, 2, \cdots, n$, where $e_{j}$ is the $j^{th}$ unit vector and $\lambda \in \mathbb{I}_n$  is a member of the index set.
 
\item Backward index: A set of backward indices for any index $\lambda$ with $\lambda_{j}>1$, is defined as $\lambda-e_{j}$.

\item Admissible set: An ordered index set $\mathbb{I}_n$ is said to be admissible if all the backward indices of any index $\lambda \in \mathbb{I}_n$ lies in $ \mathbb{I}_n$.
Mathematically, it can be represented as $
\lambda-e_{j} \in \mathbb{I}_n \quad \forall \quad \lambda \in \mathbb{I}_n, 1\leq j \leq n, \lambda_{j}>1$.
For example, an index set $\lbrace(1,1),(2,1),(1,2),(2,2)\rbrace$ is admissible while $\lbrace(1,1),(2,1),(2,2)\rbrace$ is not admissible.
    
    \item Local error indicator ($g_{\lambda}$): This indicates the error at each index. In algorithm, it is used to achieve a trade-off between accuracy and computational complexity. It is given as \cite{gerstner2003dimension}
\begin{equation}\label{eq6}
g_{\lambda}=max \left\lbrace  \psi \dfrac{\vert \boldsymbol{\bigtriangleup}_{\lambda}f \vert}{\vert\boldsymbol{\bigtriangleup}_{\textmd{I}_1}f \vert},(1- \psi)\dfrac{\varpi_{\textmd{I}_1}}{\varpi_{\lambda}}\right\rbrace  ,
\end{equation}
where $\vert \boldsymbol{\bigtriangleup}_{\lambda}f \vert$ stands for the first norm of absolute $ \boldsymbol{\bigtriangleup}_{\lambda}f$, ${\textmd{I}_1}=(1,1,\cdots,1) $ represents the first entry of the ordered index set $\mathbb{I}_n$, $\psi \in [0,1]$ is error weighting parameter and $\varpi_{\lambda}$ defines the number of function evaluations as a proxy for computational load. Later in this section, it will be understood that $\varpi_{\textmd{I}_1}$ is unity. The difference formula $\boldsymbol{\bigtriangleup}_{\lambda}f$ for a vector $\lambda=(\lambda_1,\lambda_2,\cdots,\lambda_n)$ is defined as  
\begin{equation}\label{del_dim}
\boldsymbol{\bigtriangleup}_{\lambda}f=(\bigtriangleup_{\lambda_1}\otimes\bigtriangleup_{\lambda_2}\otimes\cdots\otimes\bigtriangleup_{\lambda_n})f.
\end{equation}

\item Active index set $\mathbb{A}$: This set contains the indices whose error indicators have already been calculated and the error indicators of its forward neighbours have not been examined.

\item Old index set $\mathbb{O}$: This set holds all the other indices of $\mathbb{I}_n$ which are not included in $\mathbb{A}$.

\item Global error estimate $\mho$: It gives the sum of all $g_\lambda$ present in active index set $\mathbb{A}$.
\item $TOL$: Error tolerance value which is predefined by the user. This value decides the termination of the computation when some specific accuracy is achieved.

\end{enumerate}
\subsection{Algorithm for approximation of multivariate integrals}

The objective is to approximate the intractable integrals appeared during approximate nonlinear filtering. In SGHF, this approximation is performed with the help of tensor product of difference formulas over the indices appeared in index set $N_q^n$, as shown in \eqref{eq1}. In the proposed method, we modify the original sparse-grid construction. The selection of the index set over which the difference formula is computed and summed, is rederived in order to ignore the unwanted entries of $N_q^n$ and reduce the computational burden. To achieve this, the proposed method adopts an adaptive approach which generates fewer points along the lower nonlinear dimensions.  To this regard, the $N_q^n$ used in SGHF is replaced by an admissible index set $\mathbb{I}_n$ in ASGHF. The generation of $\mathbb{I}_n$ could be understood in a later part of this section.

The modified sparse-grid construction based on the admissible index set $\mathbb{I}_n $ can be expressed as \cite{gerstner2003dimension}
\begin{equation}\label{res_adap}
\begin{split}
\textbf{I}_{n}(f)\approx\sum_{\lambda \in \mathbb{I}_n}\boldsymbol{\bigtriangleup}_{\lambda}f^{n}(\textbf{x})=\sum_{\lambda \in \mathbb{I}_n}(\bigtriangleup_{\lambda_{1}}\otimes\cdots\otimes \bigtriangleup_{\lambda_{n}})f^{n}(\textbf{x}).
\end{split}
\end{equation}
From the above expression, it is clear that the main challenge is to generate the admissible index set ${\mathbb{I}_n}$.

The generation of the admissible index set $\mathbb{I}_n$ and functioning of the proposed algorithm can be described as follows: \\

\emph{Initialization:}

\begin{itemize}
\item First of all, two predefined controlling parameters, the error weighting parameter  and the tolerance level are set with a numeric value. Proper selection of these parameters leads to a good trade-off between the accuracy and computational burden. 

\item  The algorithm starts with the index ${\textmd{I}_1}=(1,1,\cdots,1) $, which is the first entry of index set $\mathbb{I}_n$.

\item  At the beginning, define the old index set and the active index set with a null set and $\textmd{I}_1$ respectively.

\item Compute the local error indicator for the index $\textmd{I}_1$ appeared in active index set. Also, initialize the global error estimate with an arbitrarily large value.
\end{itemize}

\emph{Processing:}

The algorithm follows the following steps:

\textit{Step 1}: Check the condition `global error estimate $>$ $TOL$'. If it is satisfied, the index with largest error indicator value is transferred from the active index set to the old index set.

\textit{Step 2}: Deduct the error indicator value of the transferred index from the last stored value of global error estimate.

\textit{Step 3}: Now, check for the forward indices of the transferred index. Add each of the forward index in active index set, if this addition does not disturb the admissibility condition.

\textit{Step 4}: As soon as all the eligible forward indices are transferred to active index set, a fresh global error estimate is calculated by adding the error indicator values of all the indices appeared in the active index set.

\textit{Step 5}: Return to step 1. If it fails to satisfy the condition discussed in step 1, we get the desired index set $\mathbb{I}_n$ by calculating the union of active index set and old index set.

\textit{Step 6}: Compute the difference formula over the index set $\mathbb{I}_n$ and add all the individual results to get the approximation of integrals, as discussed in Eq. \eqref{res_adap}.

\vspace{5mm} 
\textit{Pseudo algorithm for integral approximation}
\vspace{2mm} \\
$integrate(f)$ \\
$i:=(1,1,\cdots,1)$\\
$\mathbb{O}:=\phi$ \\
$\mathbb{A}:=\lbrace i\rbrace$\\
$r:=\boldsymbol{\bigtriangleup}_{i}f$\\
$\mho_1=g_i$\\
$initialize \ \mho \ with \ arbitrarily \ high \ value.$\\
$if \quad ( \mho_1 > TOL) \quad$\\
$select \ i\ from \ \mathbb{A} \ with \ largest \ g_{i}$\\
$\mathbb{A}:=\mathbb{A}\ - \lbrace i\rbrace$\\
$\mathbb{O}:=\mathbb{O}\cup \lbrace i \rbrace$\\
$\mho_1:=\mho_1 - g_{i}$\\
$for \ j:=1,2,\cdots,n \quad$\\
$\lambda=i+e_{k}$\\
$if \ \lambda-e_{q} \in \mathbb{O} \ for \ all \ q=1,2,\cdots,n, \ then$ \\
$\mathbb{A}:= \mathbb{A} \cup \lbrace \lambda \rbrace$ \\ $s:=\boldsymbol{\bigtriangleup}_{\lambda}f$\\ $r:=r+s$\\$\mho_1:=\mho_1+g_{\lambda}$\\ $\mho=\mho_1 $\\ $end\;if$\\ $end\;for$ \\$end\;if$\\ $return \ r$\\ where
  $\bigtriangleup_{i}f$= integral increment, $\otimes_{k=1}^{n}  \bigtriangleup_{i_{k}}f$\\ and
 $r$= computed integral value, \quad $\sum_{i\in \mathbb{O}\cup \mathbb{A}}\otimes_{k=1}^{n}\boldsymbol{\bigtriangleup}_{i_{k}}f$.

 \textit{\textbf{Note:}} In the pseudo algorithm, the parameter $\mho_1$ is actually the global error estimate $\mho$. But, it is defined separately in order to discard the stopping of algorithm before entering to the body of `$if( \mho > TOL)$'. Hence throughout the paper, no difference has been considered between these two.

\subsection{Generation of points and weights}
In the proposed method, the integral approximation is a recursive process and in each recursion, the function to be approximated is used. Hence the proposed method is model dependent. In each iteration, the function is approximated using the difference formula expressed in Eq. \eqref{del_dim} which use a set of sample points and the corresponding weights to approximate the integral. The final approximation of the integral is the sum of function approximation over all iterations. Hence, if we store all the points and corresponding weights used over different iteration, the integral can be approximated numerically using the stored points and weights. 
 
To this regard, the integrals of interest with process and measurement functions are approximated offline with a predefined tolerance, before using them in the filtering algorithm. All the points and weights used over different iterations during approximation of  the integral are stored. Then the filtering algorithm mentioned in \cite{singh2015higher} is used.

%
%
%
%
%
%
%

\subsection{Adapting to the degree of nonlinearity along different dimensions}

As discussed earlier, the proposed method puts less effort towards the dimension with lower nonlinearity \textit{i.e.} generates fewer quadrature points along those dimensions. For this purpose, it does not use a dedicated method to identify the degree of nonlinearity but the error parameter $g_{\lambda}$ helps to accomplish the objective adaptively. 

It is obvious that the difference formula $\bigtriangleup_{\lambda}$ will provide poor approximation \textit{i.e.} the absolute value of $\bigtriangleup_{\lambda}f$ will be higher if the system has higher nonlinearity. So while checking the forward indices (for next entry), a comparison between the absolute values of $\bigtriangleup_{\lambda^{i}}f$ ($\lambda^{i}$ is $i^{th}$ $\lambda$ for $i=1,2,\cdots,n$) may help to identify the dimension for which the nonlinearity is highest (the dimension with highest absolute value of $\bigtriangleup_{\lambda^{i}}f$). As soon as such dimension is identified, the next choice of index may be brought from this dimension which may help to put more effort along the dimensions with higher nonlinearity. In this regard, an error parameter $g$ is defined for each indices which makes comparison of errors computed from the difference formula.

From the above discussion, it is apparent that $g=\psi \dfrac{\vert \boldsymbol{\bigtriangleup}_{\lambda}f \vert}{\vert\boldsymbol{\bigtriangleup}_{\textmd{I}_1}f \vert}$ may be sufficient for identifying different nonlinearity in different dimensions and putting less efforts along the lower nonlinear dimensions. However, it will not help the practitioners to restrict the computational cost below a preassigned computational budget even if they are ready for a limited compromise with the accuracy. Hence, as shown in Eq. \eqref{eq6}, a second term is incorporated in the expression of $g$ which enables the practitioners to take a control over the computational cost. It justifies the earlier statement that the local error indicator $g$ also helps to access a trade-off between the accuracy and computational budget. 

\begin{rem}
If for an online implementation, the accuracy is crucial and the system is equipped to afford a high computational burden, the designer should choose a higher value of $\psi$ and vice versa. Hence, $\psi$ acts as controlling parameter which helps the algorithm to take sensible decisions when comparatively low error or high computational work is encountered.
\end{rem}

\begin{rem}
The expressions for prediction and update in approximate nonlinear filtering methods such as SGHF are widely discussed in the literature and are omitted here for brevity; the reader is referred to \cite{bar2004estimation} for example.
\end{rem}

\subsection{Selection of error weighting parameter and tolerance level}

As discussed earlier, the selection of error weighting parameter and tolerance level depends on available computational budget. To this regard, an offline implementation over the expected model will be required before going for real-life implementation. 

It is apparent from the above discussions that a high value of error weighting parameter (which lies between 0 and 1) and a low tolerance level are responsible for high accuracy but, at the same time, need a high computational time. To this regard, the practitioners may begin with a near unity value for error weighting parameter with a significantly low tolerance level. Then the required computational time for the specific selection should be compared with the available computational budget. If the required computational time remains higher, another attempt should be made with a reduced error weighting parameter or increased tolerance level or both. The same procedure should be repeated until the required computational time is not less than the available computational budget. Once this condition is achieved, the specific set of error weighting parameter and tolerance level should be chosen for online implementation.

It is to be noted here that a prior offline implementation is a common practice for many other purposes as well, like noise parameter selection, model validation \textit{etc}. Subsequently, an offline implementation constrain does not affect the reliability of the algorithm.

\subsection{Illustration}
In this subsection, the working of the algorithm is illustrated for a two dimensional nonlinear dynamic system. The dynamic behavior of the system is given as
\begin{equation}
\left[
 x_{1,k+1} \;\;
x_{2,k+1}  \right]^T=\left[ 
 e^{-x_{1,k}} \;\;
e^{-x_{2,k}^2}  \right]^T.
\end{equation} 
To evaluate the integral, initial mean and covariance is taken as $[0\;\;0]^T$ and $diag([0.4\;\;0.2])$ respectively. The error weight and tolerance are assumed as $0.725$ and $0.05$ respectively.

\begin{enumerate}
\item At starting, $\mathbb{O}=\phi$, $\mathbb{A}=\lbrace (1,1)\rbrace$, $\mho_1=50$ and $\mho=g_{(1,1)}=0.725$. $\boldsymbol{\bigtriangleup}_{(1,1)}f$  (difference formula for $(1,1)$) is calculated and result is stored in $r$, so $r=[1\;1]^T$.
\item The index with largest error indicator value is searched in $\mathbb{A}$. It is $(1,1)$, as there is single point. Its corresponding error value is subtracted from the global error, hence $\mho=0$. Then, $(1,1)$ is transferred to old index set and its forward indices are added into $\mathbb{A}$. This happens only if all the backward indices of incoming index is already present in old index set. Hence $\mathbb{O}=(1,1)$ and $\mathbb{A}=\lbrace (2,1), (1,2)\rbrace$. The difference formula for new indices are evaluated and the results are summed with the values in $r$. The error indicator value of all indices appeared in $\mathbb{A}$ are added to compute the global error value. So, $\mho_1=\mho=0.137$.
\item As the global error is still greater than the tolerance value, the algorithm will proceed further. 
\item $g$ is a set, \textit{i.e.} $g_i$ indicates the error indicator value of $i^{th}$ index of $\mathbb{A}$, then $g=\left\lbrace 0.0688\;\;0.0688\right\rbrace $.
\item  As both the indices have same value, any one can be selected, we select $(2,1)$.
\item $(2,1)$ is transferred to $\mathbb{O}$ and its forward indices $ (2,2)$ and $(3,1)$ are looked up. All the backward indices of $ (2,2)$ are not present in old index set, so only $(3,1)$ is transferred to $\mathbb{A}$. Hence $\mathbb{O}=\lbrace (1,1)\;\; (2,1)\rbrace$ and $\mathbb{A}=\lbrace (1,2)\;\; (3,1)\rbrace$.
\item The corresponding error indicator set is calculated as $g_i=\left\lbrace 0.0688 \;\;0.0344 \right\rbrace $.
\item Again the index with maximum error indicator value, \textit{i.e.} $(1,2)$, is selected from $\mathbb{A}$ and transferred to $\mathbb{O}$.
\item Similar procedure continues until we get the global error below the predefined tolerance level, 0.05. The final approximation of integral will be the value stored in $r$.
\end{enumerate}

\subsection{Advantages of ASGHF over GHF and SGHF}

\begin{enumerate}

\item The computational load of ASGHF is lower than GHF and SGHF at similar accuracy levels.

\item There are two controlling parameters in ASGHF algorithm, $viz.$ the error tolerance $TOL$ and the error weighting parameter $\psi$. These two parameters help the algorithm to find the dimensions with a higher degree of nonlinearity and refine them accordingly.

\item In GHF and SGHF, an increase in accuracy level by unity will lead to a sharp increase in the computational load. In the proposed filter, a small increase in accuracy can be acheived by varying the tolerance level or the error weighting parameter, with a proportionately small increase in the computational cost. Thus, the proposed filter gives a  comparatively smoother relation between estimation accuracy and computational cost.

\item Under assigned computational budget, the ASGHF enables the online estimator to work near the full efficiency, as the trade-off between the accuracy and computational cost can be fine-tuned, as mentioned above.

\end{enumerate}

\section{Simulation}\label{sec_simulation}

In this section, the adaptive sparse-grid Gauss-Hermite quadrature rule (used in ASGHF) is first implemented for approximation of a simple multidimensional integral, and then to two real-life nonlinear filtering problems. For all the problems its performance is compared with GHF and SGHF. During the implementation for real-life nonlinear filtering problems, a 3-point GHF and a $3^{rd}$-degree of accuracy level for SGHF (\textit{i.e.} $L = 3$) have been considered.

\subsection{Problem 1: Approximate evaluation of a multidimensional integral}

Let us assume, $\textbf{x}=[x_1$ $x_2$ $\cdots$ $x_n]^T$ be an $n$-dimensional vector, and the integral under consideration to be
\begin{equation}
I_n=\int_{-\infty}^{\infty}\sum_{i=1}^{n}x_i^{2i}d\textbf{x}.
\end{equation}
The above integral (with $n=6$) is approximated using the adaptive sparse-grid Gauss-Hermite (ASGH) quadrature rule, the sparse-grid Gauss-Hermite (SGH) quadrature rule and the Gauss-Hermite (GH) quadrature rule. The results have been compared in Table \ref{tab:0} where GH\_t represents a $t$-point GH rule, SGH\_$L$ represents SGH rule with accuracy level $L$ and  ASGH\_$\lbrace \psi,TOL \rbrace$ represents ASGH rule with error weighting parameter $\psi$ and tolerance level $TOL$. From the table, it could be concluded that the ASGH rule requires a significantly small number of sample points for achieving similar accuracy with respect to GH and SGH quadrature rules.

\begin{table}[h]

\begin{center}
\begin{tabular}{|c|c|c|}
  \hline
 Filters & \% Error & Number of sample points\\
 \hline
GH\_3 & 77.8843 & 729 \\
GH\_4 & 36.3747 & 4096 \\
GH\_5 & 7.8139 & 15625 \\
GH\_6 & 0.4784 & 46656 \\
SGH\_3 & 7.8066 & 97 \\
SGH\_4 & 0.0042 & 533 \\
ASGH\_$\lbrace$ 0.1,5 $\rbrace$ & 0.0138 & 64 \\
ASGH\_$\lbrace$ 0.4,5 $\rbrace$ & 0.0107 & 88 \\
ASGH\_$\lbrace$ 0.4,1.6 $\rbrace$ & 0.0042 & 110 \\

 \hline
\end{tabular}
\end{center}
\caption{\% error and sample point requirement for different quadrature rules}\label{tab:0}
\end{table}

\subsection{Problem 2: Estimation of multiple superimposed sinusoids}

In this problem, we estimate the amplitude and frequency of multiple superimposed sinusoids. Such problems practically appear in many fields like communication systems \cite{niedzwiecki2005estimation}, power systems \cite{reddy2009fast} \textit{etc}. 

We consider, the number of sinusoids as three, then the state variable will be $\textbf{x}=[f_1$ $f_2$ $f_3$ $a_1$ $a_2$ $a_3]^T$, where $f_i$ and $a_i$ are the frequency and amplitude of $i^{th}$ sinusoid. The  discretized process model is 
\begin{equation}
\textbf{x}_k=\texttt{I}_6\textbf{x}_{k-1}+w_k,
\end{equation}
where $\texttt{I}_6$ is a six dimensional unit matrix and $w_k$ is process noise normally distributed with zero mean and covariance $Q=diag([\sigma_f^2\;\;\sigma_f^2\;\;\sigma_f^2\;\;\sigma_a^2\;\;\sigma_a^2\;\;\sigma_a^2])$ with $\sigma_f$ and $\sigma_a$ being the standard deviations for frequency and amplitude.

The measurement equation is \cite{closas2012multiple}
\begin{equation*}\label{me}
y_k=\left[ \begin{array}{c}
 \sum_{j=1}^{3}a_{j,k}cos(2\pi f_{j,k}kT) \\
\sum_{j=1}^{3}a_{j,k}sin(2\pi f_{j,k}kT)  \end{array} \right] +v_k,
\end{equation*}
where $v_k$ is Gaussian noise with zero mean and covariance $R=diag([\sigma_n^2\;\;\sigma_n^2])$ with $\sigma_n$ being the standard deviation for measurement noise. $T$ is the sampling time which is considered as 
0.1667 $ms$.

The initial truth and estimates are considered as $[200$ $1000$ $2000$ $5$ $4$ $3]^T$ and $[150$ $900$ $1800$ $4$ $4$ $2]^T$ respectively. Varying the initial error covariance and noise covariances, we consider two different scenarios as

\textit{scenario 1:}

$\sigma_f^2=151 \mu Hz^2/ms^2$,
 $\sigma_a^2=80 \mu V^2/ms^2$,
  $\sigma_n^2=0.09 V^2$, and
$P_{0|0}=diag([20^2$ $ 20^2$ $ 20^2$ $ 0.05$ $ 0.05$ $ 0.05])$.

\textit{scenario 2:}

$\sigma_f^2=300 \mu Hz^2/ms^2$,
 $\sigma_a^2=160 \mu V^2/ms^2$,
  $\sigma_n^2=0.18 V^2$, and
$P_{0|0}=diag([50^2$ $ 50^2$ $ 50^2$ $ 0.5 $ $0.5 $ $0.5])$.

 For ASGHF, the simulation is performed by considering the error weighting parameters as 0.6 and 0.5, while tolerance as 0.53 and 0.6655 for process and measurement equations respectively.
 The states are estimated for 500 steps and the results are averaged over 2000 Monte Carlo runs. At each step, a combined error parameter ($ERR$) is evaluated for frequency and amplitude, which is defined as
\begin{equation}
ERR_k=\sqrt{\frac{MSE_{1,k}+MSE_{2,k}+MSE_{3,k}}{3}},
\end{equation}
where, for $M$ number of Monte Carlo runs, $MSE_{i,k}$ is
\begin{equation}
MSE_{i,k}=\frac{1}{M}\sum_{j=1}^{M}(\textbf{x}_{i,k,j}-\hat{\textbf{x}}_{i,k,j})^2.
\end{equation}

%

%

\begin{table}[h]

\begin{center}
\begin{tabular}{|l|c|}
  \hline
 Filters & Relative comp. time (Prob. 1)\\
 \hline
GHF & 1 \\
SGHF & 0.17 \\
ASGHF & 0.056 \\
 \hline
\end{tabular}
\end{center}
\caption{Relative computational time for various filters}\label{tab:1}
\end{table}

\begin{center}
\begin{figure}[!htbp]
\includegraphics[width=6in,height=5in]{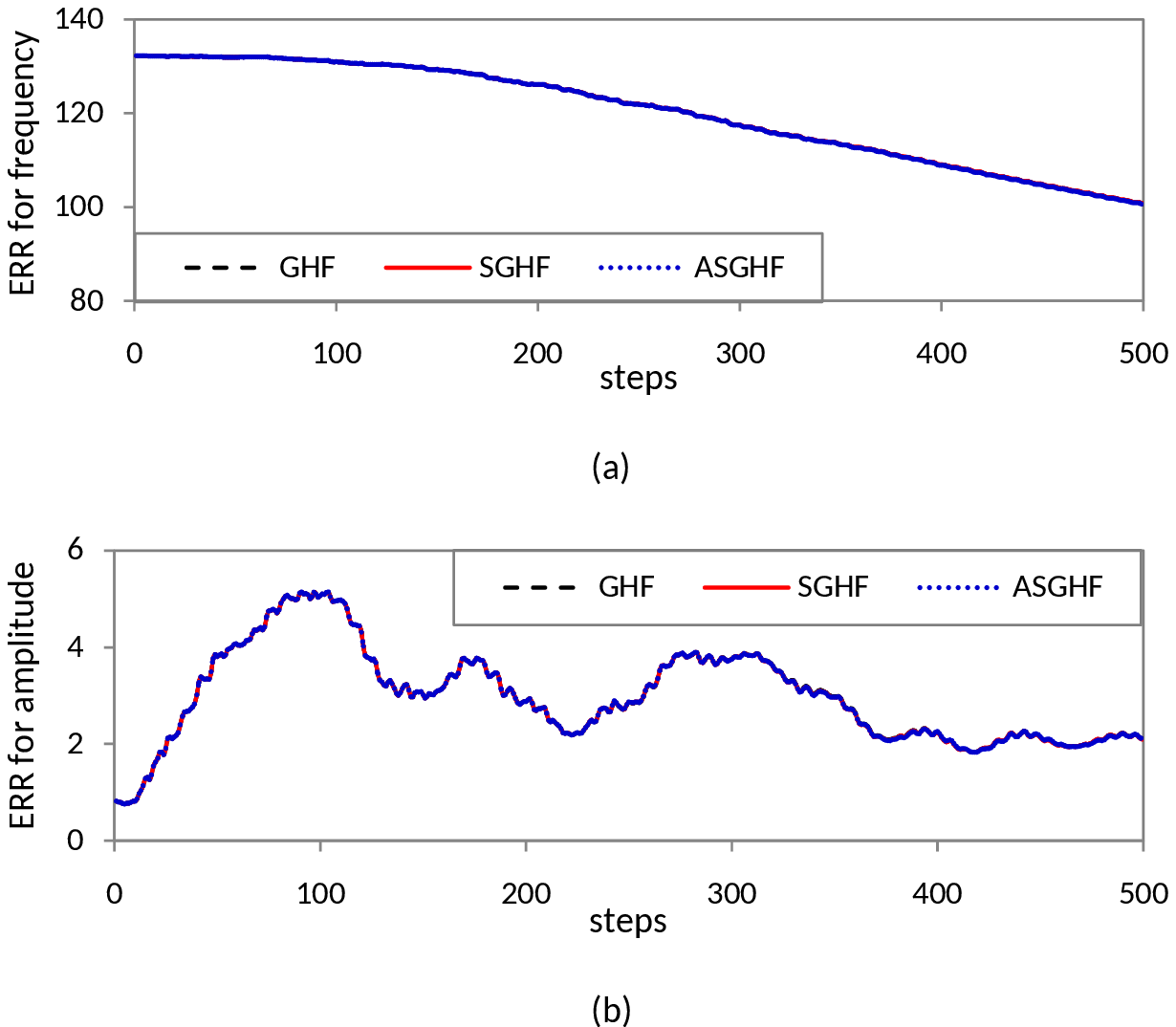}
\caption{\textit{Problem 1:} ERR plot for $1^{st}$ scenario- (a) frequency in Hz (b) amplitude in $volt$.}
\label{fig:scene1}
\end{figure}
\end{center}

\begin{center}
\begin{figure}[!htbp]
\includegraphics[width=6in,height=5in]{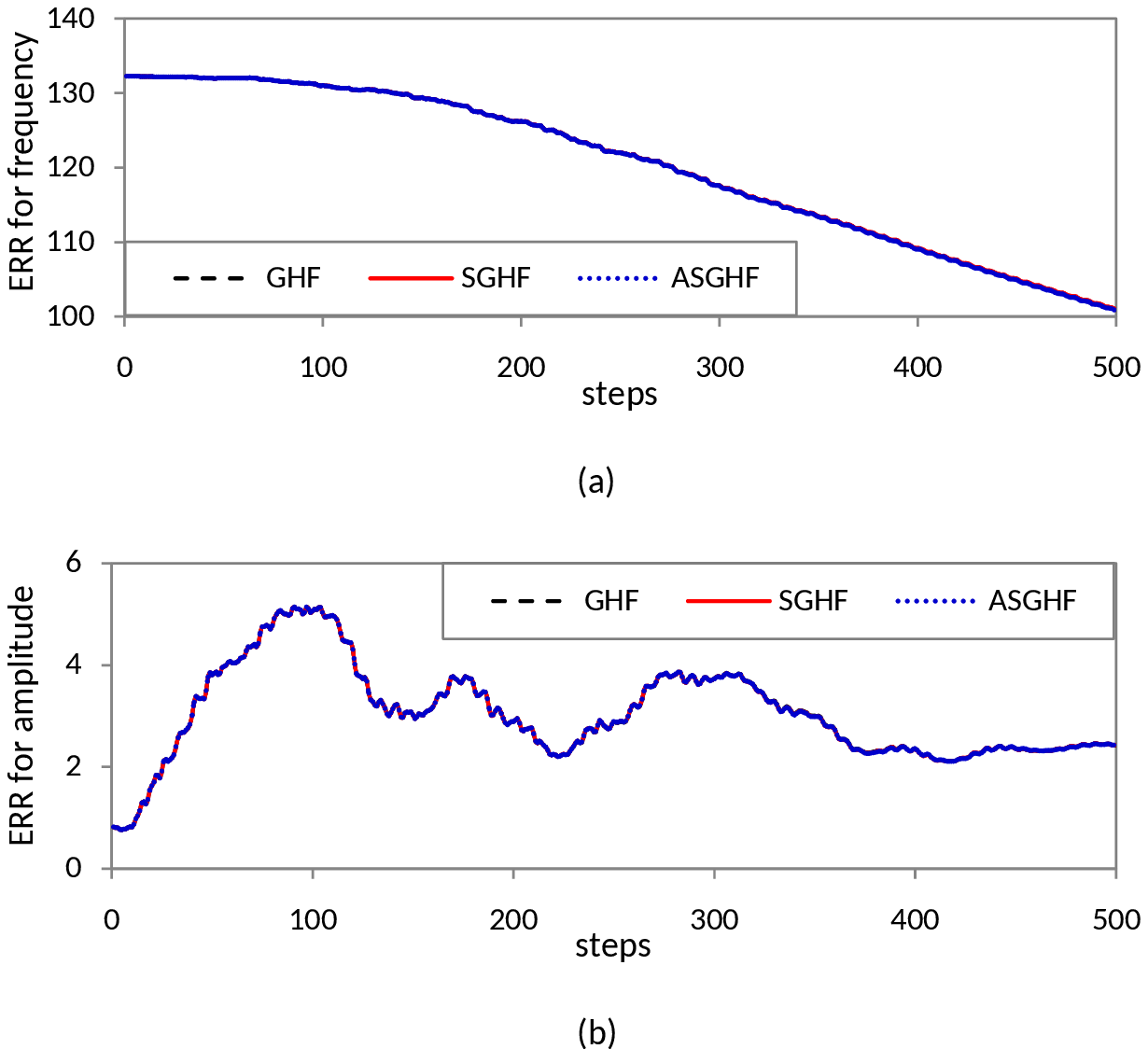}
\caption{\textit{Problem 1:} ERR plot for $2^{nd}$ scenario- (a) frequency in Hz (b) amplitude in $volt$.}
\label{fig:scene2}
\end{figure}
\end{center}

The $ERR$ for frequency and amplitude are plotted for two different scenarios in Fig. \ref{fig:scene1} and Fig. \ref{fig:scene2} using the proposed ASGHF, SGHF and GHF. The $ERR$ is similar for all the filters and hence it could be concluded that the accuracy of the proposed algorithm is similar to the conventional GHF and SGHF. On the other hand, from the Table \ref{tab:1}, it could be concluded that the computational burden for the proposed ASGHF is almost 3 times lower than the SGHF and 18 times lower than the conventional GHF. 

\subsection{Problem 3: Maneuvering target tracking}
The second problem is a tracking problem of a target following coordinated turn model \cite{bar2004estimation}. The discretized target dynamics can be represented as
\begin{equation}
\textbf{x}_{k+1}=F_{k}\textbf{x}_{k}+w_{k},
\end{equation}
where $\textbf{x}=[\texttt{x} \ \dot{\texttt{x}} \ \texttt{y} \ \dot{\texttt{y}} \ \omega]^{T}$ with \texttt{x} and \texttt{y} being the positions in $x$ and $y$ directions respectively, $\omega$ is the angular turn rate and
 \[F_{k}= \left[ \begin{array}{ccccc}
1 & \dfrac{\sin(\omega_{k}T)}{\omega_{k}} & 0 & -\dfrac{1-\cos(\omega_{k}T)}{\omega_{k}} & 0 \\
0 & \cos(\omega_{k}T) & 0 & -\sin(\omega_{k}T) & 0 \\
0 & \dfrac{1-\cos(\omega_{k}T)}{\omega_{k}} & 1 & \dfrac{\sin(\omega_{k}T)}{\omega_{k}} & 0 \\
0 & \sin(\omega_{k}T) & 0 & \cos(\omega_{k}T) & 0 \\
0 & 0 & 0 & 0 & 1 \end{array} \right].\]

The nonlinear measurement equation can be described in general as
\begin{equation}
y_{k}=\gamma(\textbf{x}_{k})+v_{k}.
\end{equation}
Here, we assume that both the range and bearing angle are available from measurements and hence, the measurement equation can be written as
\begin{equation*}\label{me}
y_{k}=\left[ 
 \sqrt{\texttt{x}_k^2+\texttt{y}_k^2} \;\;\;\; 
 atan2(\texttt{y}_k,\texttt{x}_k)  \right]^T \\+v_k,
\end{equation*}
where $atan2$ is the four quadrant inverse tangent function. $w_{k}$ and $v_{k}$ are considered to be white Gaussian noise with zero mean and covariances $Q$ and $R$ respectively. The process noise covariance is
\begin{equation*}
Q=q\left[ \begin{array}{ccccc}
 \dfrac{T^3}{3} &  \dfrac{T^2}{2} & 0 & 0 & 0 \\
\dfrac{T^2}{2} & T & 0 & 0 &  0 \\
0 & 0 & \dfrac{T^3}{3} & \dfrac{T^2}{2} & 0 \\
  0 & 0 & \dfrac{T^2}{2} & T &  0 \\
  0 & 0 & 0 & 0 & 0.009T  \end{array} \right], 
\end{equation*}
where $q=0.1$ is a given constant and $T$=0.5 seconds is the sampling time. $R=\text{diag}([\sigma_r^2 \,\, \sigma_t^2])$, where $\sigma_r=120\text{m}$ and $\sigma_t=\sqrt{70}\text{mrad}$.

The initial truth value is considered as $\textbf{x}_0=[1000\text{m} \,\,\, 30\text{m/s}\,\,\,$ $1000\text{m}\,\,\, 0\text{m/s}\,\,\, \omega^{\circ}/s]^T$, while the initial covariance is  $P_{0|0}=\text{diag}([200\text{m}^2\,\,\, 20\text{m}^2/\text{s}^2\,\,\, 200\text{m}^2\,\,\, 20\text{m}^2/\text{s}^2\,\,\, 100\text{mrad}^2/\text{s}^2]) $. The initial estimate is considered to be normally distributed with mean $\textbf{x}_0$ and covariance $P_{0|0}$.
 For ASGHF, we consider two different scenarios by selecting different set of predefined parameters as shown in Table \ref{tab:a}.

\begin{table}
\begin{center}
\begin{tabular}{|c|c|c|c|c|}
  \hline
Model &\multicolumn{2}{c|}{scenario 1} &\multicolumn{2}{c|}{scenario 2}\\
 \cline{2-5}
 &\textit{$\psi$}&\textit{TOL} &\textit{$\psi$}&\textit{TOL}\\
 \hline
Process model & 0.55 & 0.5 & 0.525 & 0.5\\
 \hline
Measurement model & 0.6 & 0.48 & 0.6 & 0.48\\
 \hline
\end{tabular}
\end{center}
\caption{Two different scenarios for Problem 3}\label{tab:a}
\end{table}

\begin{center}
\begin{figure}[!htbp]
\includegraphics[width=6in,height=5in]{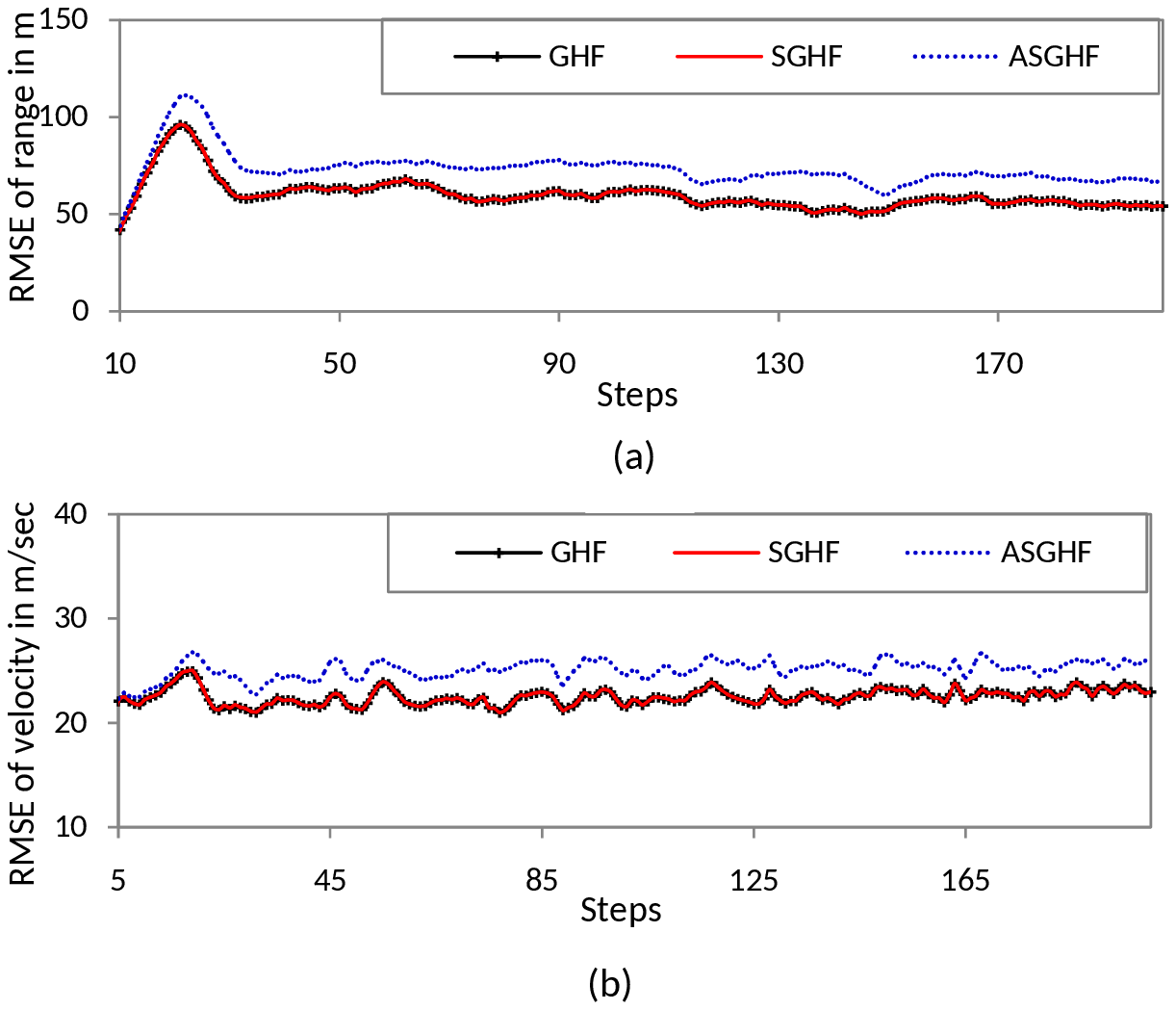}
\caption{\textit{Problem 2:} RMSE plots for $\omega=3^{\circ}$ under $1^{st}$ scenario- (a) range in $m$ (b) velocity in $m/s$.}
\label{fig:1m}
\end{figure}
\end{center}

\begin{center}
\begin{figure}[!htbp]
\includegraphics[width=6in,height=5in]{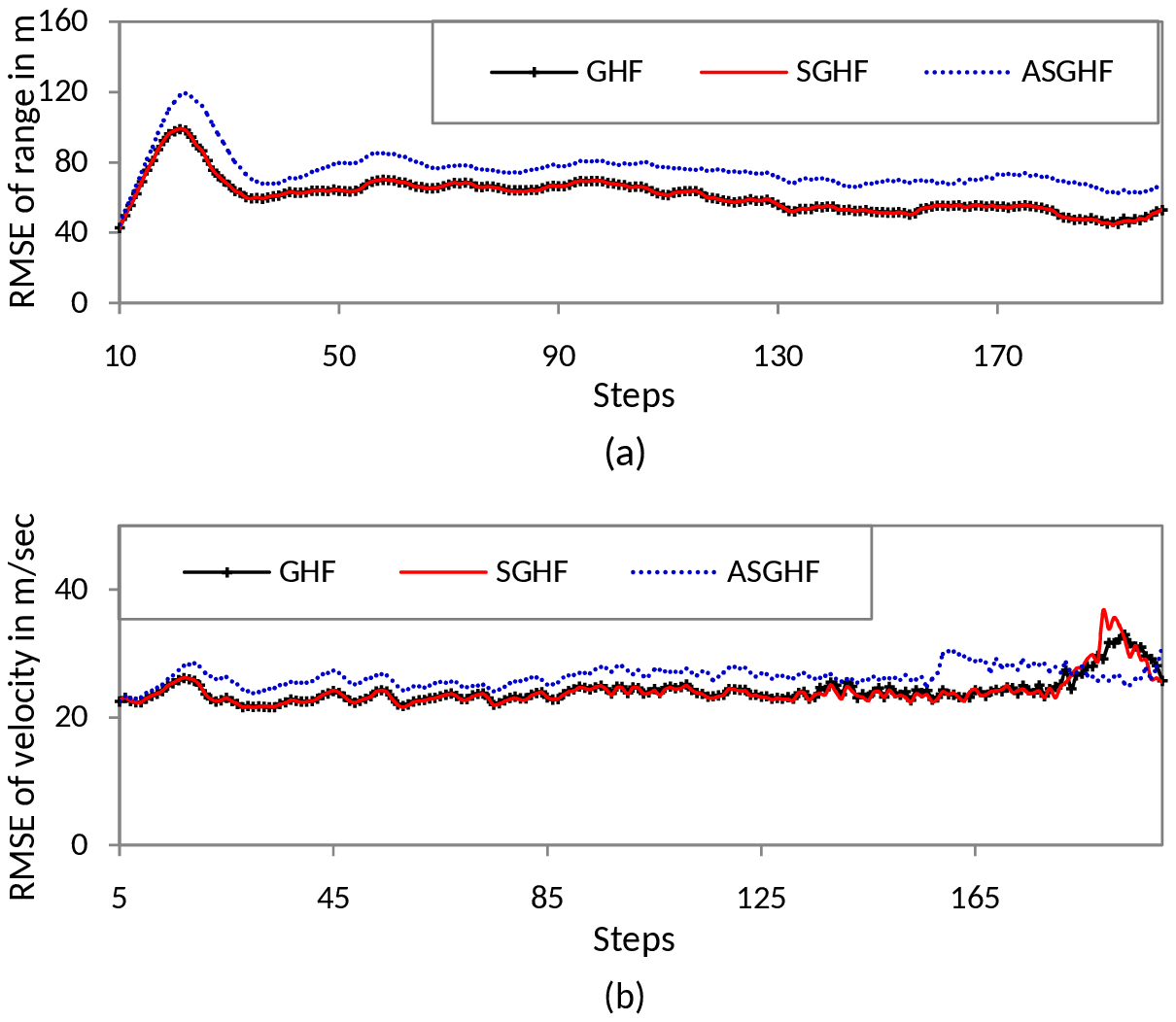}
\caption{\textit{Problem 2:} RMSE plots for $\omega=4.5^{\circ}$ under $1^{st}$ scenario-  (a) range in $m$ (b) velocity in $m/s$}
\label{fig:2m}
\end{figure}
\end{center}

\begin{center}
\begin{figure}[!htbp]
\includegraphics[width=6in,height=5in]{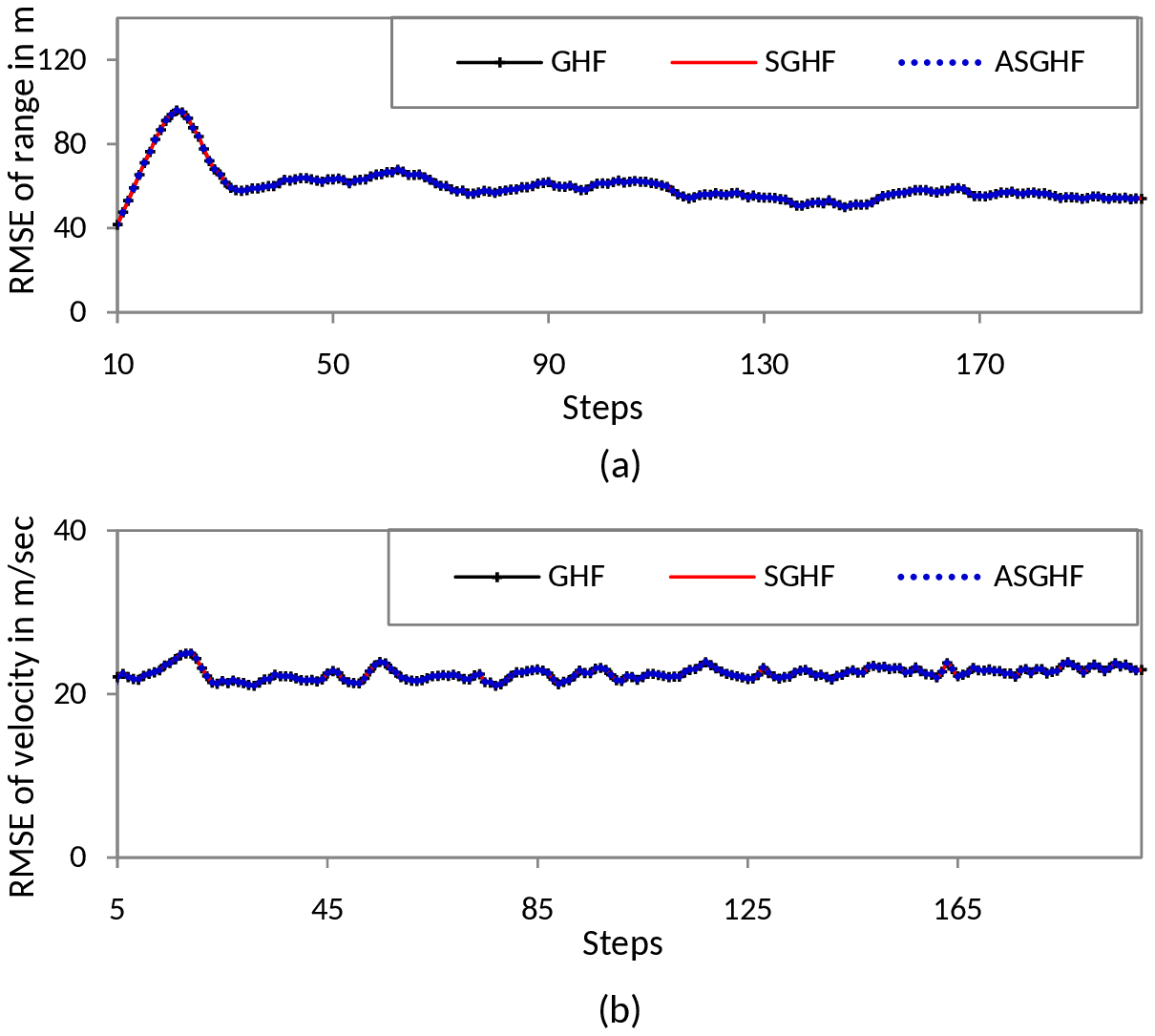}
\caption{\textit{Problem 2:} RMSE plots for $\omega=3^{\circ}$ under $2^{nd}$ scenario- (a) range in $m$ (b) velocity in $m/s$.}
\label{fig:4m}
\end{figure}
\end{center}

\begin{center}
\begin{figure}[!htbp]
\includegraphics[width=6in,height=5in]{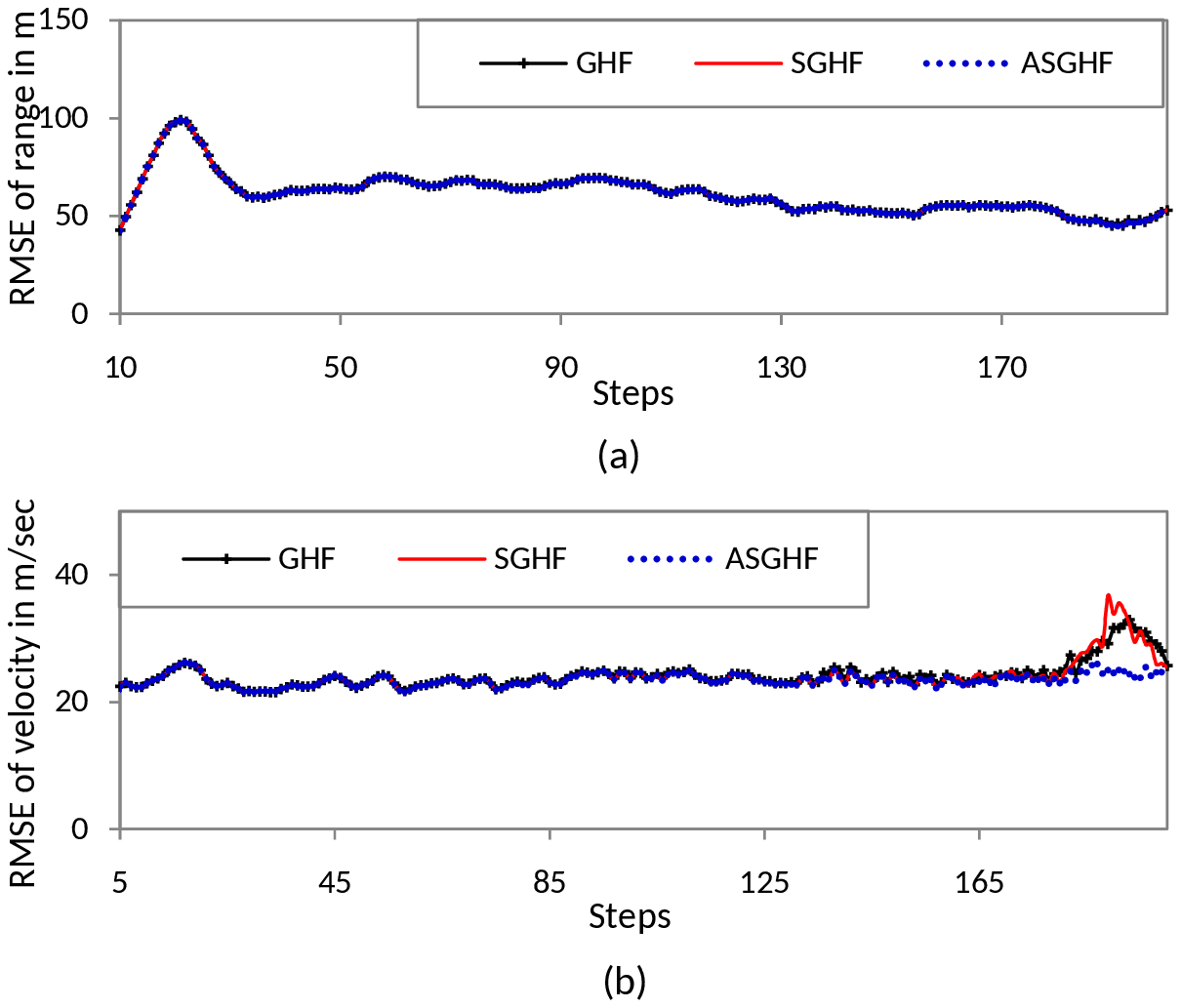}
\caption{\textit{Problem 2:} RMSE plots for $\omega=4.5^{\circ}$ under $2^{nd}$ scenario-  (a) range in $m$ (b) velocity in $m/s$}
\label{fig:5m}
\end{figure}
\end{center}

The motivation of considering two parametric scenarios is to study the experimental behavior of the proposed method over the accuracy and the computational burden by varying these predefined parameters. To this regard, the simulation is performed for 100 seconds and the results are obtained in terms of root mean square error (RMSE) of the range and the velocity for 500 independent Monte Carlo runs. The performance of the proposed method is studied and compared for varying turn rate. The RMSEs are plotted for $\omega=3^{\circ}/sec$ and $\omega=4.5^{\circ}/sec$ in Fig. \ref{fig:1m} and Fig. \ref{fig:2m} respectively for scenario 1 and in Fig. \ref{fig:4m} and Fig. \ref{fig:5m} respectively for scenario 2. At the same time, the relative computational burdens are listed in Table \ref{tab:2}.

\begin{table}[h]

\begin{center}
\begin{tabular}{|c|c|c|}
  \hline
 Filters &\multicolumn{2}{l|}{Relative comp. time (Prob. 2)}\\
 \cline{2-3}
 &\textit{scenario 1}&\textit{scenario 2}\\
 \hline
GHF & 1 & 1\\
SGHF & 0.36 & 0.36 \\
ASGHF & 0.12 & 0.28 \\
 \hline
\end{tabular}
\end{center}
\caption{Relative computational time for various filters}\label{tab:2}
\end{table}

 Under the first scenario, from the Table \ref{tab:2} and Fig. \ref{fig:1m} and \ref{fig:2m}, it could be concluded that the computational burden for the proposed method is around $1/8^{th}$ and  $1/3^{rd}$ times lower than the conventional GHF and SGHF respectively, but it lags behind in terms of accuracy. In order to achieve similar accuracy, the error weighting parameter is tuned to a value as presented in scenario 2 (Table \ref{tab:a}). The computational load increases in this scenario but it still remains less than the GHF and SGHF, while a similar accuracy to these filters could be obtained. It is to be noted that the different computational times are obtained on a personal computer with 64-bit operating system, 4 GB RAM and 3.33 GHz clock speed, on a MATLAB version 2010b.

 
From the tables, it could be concluded that the improvement in computational efficiency using the proposed method is not similar for both the problems. This dissimilarity is because of the different dimension and the different degree of nonlinearity for process and measurement models. As any of these two factors increases, the degree of improvement in computational efficiency rises.

%



%

\section{Discussions and conclusions}\label{sec_conclusion}

The Gauss-Hermite quadrature rule based filters, namely the GHF and the SGHF are among the most accurate nonlinear filtering approximations available in literature. However, these filters are often not fit for real-life on-board implementation because of their high computational burden. Apart from high computational burden, another serious disadvantage with these filters is that a unit increase in the accuracy level or the number of univariate quadrature points leads to an exponential rise in the computational burden. Hence, the online estimators mostly work much below their full efficiency.

To overcome these disadvantages, this paper proposes a new Gauss-Hermite quadrature rule based filtering technique, named as adaptive sparse-grid Gauss-Hermite filter (ASGHF). It could reduce the computational burden without compromising with the estimation accuracy. Moreover, the presence of two predefined control parameters, namely the tolerance level and the error weighting parameter, help in obtaining a better trade off between the accuracy and computational load. 

The proposed filter as well as the GHF and SGHF are implemented to solve two different  state estimation problems. Simulation results show that the accuracy of ASGHF is similar to the GHF and SGHF while the computational burden is considerably less. Hence, the proposed method has potential to replace the existing filters for real-life applications.




\newpage

\newpage

\newpage

\newpage

\newpage

\newpage

\end{document}